\documentclass[prd,nofootinbib,preprint]{revtex4}
\usepackage[T1]{fontenc}
\usepackage[utf8]{inputenc} 
\usepackage{amsmath,amssymb}
\usepackage{epsfig}
\usepackage{graphicx,array}
\usepackage[usenames,dvipsnames]{color}
\usepackage{slashed}
\usepackage{comment}
\usepackage{booktabs}

\usepackage[colorlinks,citecolor=blue]{hyperref}
\usepackage{pdfpages}
\usepackage{color}
\usepackage{subfigure}

\usepackage{multirow}

\begin{document}
	
	\title{Cosmological signature and light Dark Matter in Dirac $L_\mu-L_\tau$ model}
	
	\author{Pritam Das}
	\email{prtmdas9@gmail.com}
	\affiliation{Department of Physics, Salbari College, Baksa, Assam-781318, India}

	\begin{abstract}
		We revisit an anomaly-free extension of the Standard Model (SM) $viz.$ gauged ${L_\mu-L_\tau}$ model in the Dirac framework, where the local $U(1)_{L_\mu-L_\tau}$ symmetry breaks and gives rise to a new gauge boson $Z'$ and corresponding gauge coupling $g_{\mu\tau}$. Three additional heavy vector-like fermions, three light right-handed neutrinos and two heavy singlet scalars are added to complete the model framework for Dirac neutrinos. Another singlet vector-like fermion is added with a new gauge charge, which serves as a viable DM candidate, and the correct relic abundance is obtained via the resonance effect. The resonance effect triggers early kinetic decoupling of the DM candidate, which significantly influences the cosmological sector of this study. The parameter space is considered after satisfying the current bounds on $M_{Z'}$ and the gauge coupling $g_{\mu\tau}$. The influence of dark radiations coming from the additional light degrees of freedoms are studied in connection with the dark matter. After imposing the relevant theoretical and experimental constraints, the viable parameter space is found to be significantly restricted while remaining accessible to ongoing and near-future experiments. The framework provides a consistent setting in which neutrino masses, dark matter, and cosmological observables can be studied simultaneously. In particular, we identify viable regions of the parameter space that satisfy the constraints from neutrino masses, dark matter relic abundance, and cosmological observables.
	\end{abstract}
	
	\maketitle
	
	\section{Introduction}
	
	The experimental discovery of neutrino mass demands beyond the Standard Model (BSM) frameworks. However, the exact nature of them is still an enigma as to whether they are {\it Dirac} or {\it Majorana }? The essential difference lies in the fact that in the Dirac scenario, there are a minimum of two more light degrees of freedom (DOF) known as right-handed neutrinos ($\nu_R$) than in the Majorana scenarios. The presence of additional light DOF contributes to the effective relativistic DOF, $\Delta N_{\rm eff}$. The SM prediction $N_{\rm eff}=3.045$ \cite{Mangano:2005cc, Grohs:2015tfy, deSalas:2016ztq},  is consistent with the current Planck 2018 measurement, $N_{\rm eff}=2.99\pm0.17$ \cite{Planck:2018vyg}. A similar bound also exists from big bang nucleosynthesis (BBN) $2.3 < N_{\rm eff} <3.4$ at $95\%$ CL \cite{Cyburt:2015mya}. Future experiments like CMB Stage IV (CMB-S4) is expected to reach an unprecedented sensitivity of $\Delta { N}_{\rm eff}={N}_{\rm eff}-{ N}^{\rm SM}_{\rm eff}
	= \pm0.06$ \cite{Abazajian:2019eic}, taking it closer to the SM prediction. Enhancements of $\Delta N_{\rm eff}$ in Dirac neutrino models have been studied in several recent works \cite{Nanda:2019nqy,Li:2022yna, Abazajian:2019oqj, FileviezPerez:2019cyn, Han:2020oet, Luo:2020sho, Adshead:2020ekg,Luo:2020fdt, Mahanta:2021plx, Biswas:2021kio, Borah:2022obi,Borah:2020boy,Biswas:2022fga, Biswas:2022vkq, Borah:2022enh}
	
	Due to the superlight nature and weak interaction strength, $\nu_R$ decouples early, influencing the effective relativistic degrees of freedom. If $\nu_R$ are in thermal equilibrium and they decouple at temperature $T_{\nu_R}^{dec}$, then the change in the effective numbers of neutrino species with the standard value $N_{\rm eff}^{\rm SM}$ can be expressed as \cite{Mangano:2005cc, Luo:2020sho}:
	\begin{eqnarray}
		\Delta N_{\rm eff}=N_{\rm eff}-N_{\rm eff}^{SM}=N_{\nu_R}\Big[\frac{g_*(T_{\nu_L}^{dec})}{g_*(T_{\nu_R}^{dec})}\Big]^{4/3},
	\end{eqnarray}  
	where, $N_{\nu_R}$ is number of RH neutrino species.

	The anomalous magnetic moments of leptons, especially those of the muon and electron, are also important for the SM. It has been a long-term test of the SM and has put strict limits on BSM theories. The anomalous magnetic moment of leptons has garnered significant attention in the literature, as it contests the phenomenological efficacy of the Standard Model. 
	The Fermilab Muon $g-2$ experiment has released its final and most precise measurement of the muon anomalous magnetic moment, with a world-average value
	$a_\mu^{\rm Exp}=116592071.5(14.5)\times 10^{-11}$ at $\sim124$ ppb precision \cite{Muong-2:2025xyk}, which sets a new benchmark for this quantity. This result is consistent, within current theoretical uncertainties, with recent Standard Model evaluations such as $a^{\rm SM}_\mu = 116 592 033(62)\times 10^{-11}$ \cite{Aliberti:2025beg}, while further improvements in the precision of the Standard Model prediction are needed to draw a final conclusion, please see for recent references \cite{ DiLuzio:2024sps, Hertzog:2025ssc,Hoferichter:2025yih, Kuberski:2024bcj}.

	In recent years, several works have focused on incorporating the muon $(g-2)$ from the model-building prospects. For example, see \cite{Arcadi:2021cwg,Zhu:2021vlz,Han:2021gfu, Baum:2021qzx, Bai:2021bau, Das:2021zea, Lu:2021vcp} for minimal dark matter (DM) motivated scenarios, \cite{Ge:2021cjz, Brdar:2021pla, Buen-Abad:2021fwq} for axion-like particle (ALP) motivated scenarios, \cite{Zu:2021odn, Amaral:2021rzw} for gauged lepton flavour models \cite{Heeck:2014zfa,Endo:2021zal, Ahmed:2021htr, Greljo:2022dwn, Abdughani:2021pdc, VanBeekveld:2021tgn, Cox:2021gqq, Wang:2021bcx, Gu:2021mjd, Cao:2021tuh, Yin:2021mls, Han:2021ify, Aboubrahim:2021rwz, Yang:2021duj,Chakraborti:2021bmv,  Ferreira:2021gke, Wang:2021fkn, Li:2021poy, Cadeddu:2021dqx, Calibbi:2021qto, Chen:2021vzk, Escribano:2021css, Chun:2021dwx, Arcadi:2021yyr, Chen:2021jok, Nomura:2021oeu,Mahapatra:2023zhi,Borah:2025wcc, De:2025hay,De:2024tvj, Figueroa:2024tmn}. 
	In leptonic model-building sectors, models in which the mediators couple exclusively to the second and third generations of leptons and are anomaly-free can exhibit attractive experimental features. Here we consider the widespread and minimal anomaly-free model based on the gauged $L_{\mu}-L_{\tau}$ symmetry \cite{He:1990pn, He:1991qd}. The gauged $U(1)_{L_\mu-L_\tau}$ introduces an additional gauge boson $Z'$ and this new gauge boson can give an additional contribution to $(g-2)_\mu$ via one-loop level. To keep the model consistent with current $(g-2)_\mu$, we stick to a parameter space for $Z'$ mass and the related gauge coupling $g_{\mu\tau}$ in the allowed region.
	To construct the model framework, we have introduced three heavy vector-like fermions ($N_{e,\mu,\tau}$), three right-handed Dirac neutrinos ($\nu_{(e,\mu,\tau)R}$) and two heavy scalar singlets ($\Phi_1,\Phi_2$). These fields are associated with neutrino mass generation, where light neutrino masses are generated via the Dirac SeeSaw (DSS) mechanism \cite{Borboruah:2024lli, Borah:2022obi, Chen:2022bjb, Borah:2017dmk}.  We have restricted the Lagrangian to terms that do not violate lepton number by using the $U(1)_{L_\mu-L_\tau}$ symmetry.
	
	Nevertheless, the possibility of a rich phenomenology arising from the dark sector remains open. There is strong observational evidence for dark matter (DM) from various astrophysical and cosmological observations, such as WMAP \cite{WMAP:2012fli, WMAP:2003ivt} and Planck \cite{Planck:2018vyg}. The rigorous search for heavy DM masses ranging from GeV still continues, yet there is no concrete signal of its existence \cite{LZ:2022ufs}. It is theoretically plausible that new particles exist with masses far below the electroweak scale and their interactions with ordinary matter are sufficiently weak. Such feebly interacting states fall into the "dark sector" and naturally arise in many extensions of the Standard Model. The search for light DM masses in the MeV–GeV range has gained sufficient attention in recent years due to high-intensity fixed-target experiments and several low-threshold direct search experiments, which provide the most sensitive probes \cite{Beacham:2019nyx, Ilten:2022lfq, Essig:2022dfa, ParticleDataGroup:2024cfk, Du:2020ldo}.
	In connection with the rest of the phenomenology, we have introduced a singlet vector-like fermion, $\psi$, charged under the new gauge group and it behaves as a viable dark matter candidate in our study. The light $Z'$ mediated annihilation processes allow us to consider a viable dark matter parameter region with mass in the sub-GeV scale. Such light particles (mass$<\mathcal{O}(100)$ MeV) can also decay or annihilate to the SM neutrinos and contribute to ${\Delta N_{\rm eff}}$ \cite{Baumann:2017gkg, Ghosh:2024cxi, Blennow:2012de, Yeh:2022heq, Breitbach:2018ddu}.

	This work is organized as follows: after giving a brief model description in section \ref{sec2}, we explore a Dirac neutrino mass model and muon ($g-2$) constraints under the same section. Cosmological consequences are explored in section \ref{sec3}, where the fate of the RH neutrinos, dark matter, { early kinetic decoupling due to resonance}  and contributions to $\Delta N_{\rm eff}$ are explored in detail in subsequent subsections. Finally, we conclude our work in section \ref{sec4}.   
	
	\section{Model Framework}\label{sec2}
	\subsection{ Dirac $L_{\mu}-L_{\tau}$ Model}
	The SM fermion content with their gauge charges under $SU(3)_c \times SU(2)_L \times U(1)_Y \times U(1)_{L_{\mu}-L_{\tau}}$ gauge symmetry are denoted as follows.
	
	$$ q_L=\begin{pmatrix}u_{L}\\
		d_{L}\end{pmatrix} \sim (3, 2, \frac{1}{6}, 0), \; u_R (d_R) \sim (3, 1, \frac{2}{3} (-\frac{1}{3}), 0)$$
	$$L_e=\begin{pmatrix}\nu_{e}\\
		e_{L}\end{pmatrix} \sim (1, 2, -\frac{1}{2}, 0), \; e_R \sim (1, 1, -1, 0) $$
	$$L_{\mu}=\begin{pmatrix}\nu_{\mu}\\
		\mu_{L}\end{pmatrix} \sim (1, 2, -\frac{1}{2}, 1), \; \mu_R \sim (1, 1, -1, 1) $$
	$$L_{\tau}=\begin{pmatrix}\nu_{\tau}\\
		\tau_{L}\end{pmatrix} \sim (1, 2, -\frac{1}{2}, -1), \;  \tau_R \sim (1, 1, -1, -1)$$  \\
	
	Since we want to realise Dirac neutrinos in this model, we first introduce three right handed neutrinos $\nu_{eR}, \nu_{\mu R}, \nu_{\tau R}$ singlets under the SM gauge symmetry and with $L_{\mu}-L_{\tau}$ charges $0, n_{X}, -n_{X}$ respectively. While this combination remains anomaly-free for any $n_{X}$, the importance of its value will become clear later. If $n_{X}=1$, one can have a diagonal Dirac neutrino mass matrix with neutrinos coupling to the SM Higgs. Apart from the fine-tuned Yukawa required to generate sub-eV Dirac neutrino mass, such a diagonal neutrino mass matrix is also inconsistent with neutrino oscillation data \cite{Zyla:2020zbs}, given the fact that the charged lepton mass matrix is also diagonal.
	
	\begin{table}[h!]
		\begin{center}
			\small
			\begin{tabular}{||@{\hspace{0cm}}c@{\hspace{0cm}}|@{\hspace{0cm}}c@{\hspace{0cm}}|@{\hspace{0cm}}c@{\hspace{0cm}}|@{\hspace{0cm}}c@{\hspace{0cm}}||}
				\hline
				\hline
				\begin{tabular}{c}
					{\bf ~~~~ Gauge~~~~}\\
					{\bf ~~~~Group~~~~}\\ 
					\hline
					
					$SU(2)_{L}$\\ 
					\hline
					$U(1)_{Y}$\\ 
					\hline
					$U(1)_{L_\mu-L_\tau}$\\ 
				\end{tabular}
				&
				&
				\begin{tabular}{c|c}
					\multicolumn{2}{c}{\bf Fermion}\\
					\hline
					~~~$N_{(e, \mu, \tau)L} $~~~& ~~~$\nu_{eR}, \nu_{\mu R}, \nu_{\tau R}$~~~ \\
					\hline
					$1$&$1$\\
					\hline
					$0$&$0$\\
					\hline
					$0, 1, -1$&$0, n_{X}, -n_{X}$\\
				\end{tabular}
				&
				\begin{tabular}{c|c}
					\multicolumn{2}{c}{\bf Scalar}\\
					\hline
					~~~$\Phi_{1}$~~~& $\Phi_{2}$ \\
					\hline
					$1$ & $1$\\
					\hline
					$0$ & $0$ \\
					\hline
					$1-n_{X}$ & $1$\\
				\end{tabular}\\
				\hline
				\hline
			\end{tabular}
			\caption{New particles and their corresponding gauge charges relevant for type I Dirac seesaw.}
			\label{tab1}
		\end{center}    
	\end{table}
	We have added three vector-like fermions ($N_e,N_\mu,N_\tau$) and two singlet scalars $\Phi_1$ and $\Phi_2$ with VEVs $v_1$ and $v_2$ respectively. Respective charges for the particle content are shown in Table \ref{tab1}. Apart from the Dirac neutrinos, the gauge group $U(1)_{L_\mu-L\tau}$ introduces one additional particle into the list: the new gauge boson $Z'$. The interaction Lagrangian with the new gauge boson $Z'$ is given by:
	\begin{eqnarray}\label{zplag} 
		\nonumber \mathcal{L}_{int}&\supset& i \overline{N_{eL}}\gamma^\alpha D_\alpha N_{eL}+i \overline{N_{\mu L}}\gamma^\alpha D_\alpha N_{\mu L}+i \overline{N_{\tau L}}\gamma^\alpha D_\alpha N_{\tau L}\\
		&&\nonumber +i \overline{\nu_{eR}}\gamma^\alpha D_\alpha \nu_{eR}+i \overline{\nu_{\mu R}}\gamma^\alpha D_\alpha \nu_{\mu R}+i \overline{\nu_{\tau R}}\gamma^\alpha D_\alpha \nu_{\tau R}\\
		&&\nonumber+(D_\alpha\Phi_1)^\dagger(D^\alpha\Phi_1)+(D_\alpha\Phi_2)^\dagger(D^\alpha\Phi_2)-\frac{1}{4}Z'^{\alpha\beta}Z'_{\alpha\beta}\\&&+\frac{\epsilon}{4}Z'^{\alpha\beta}F_{\alpha\beta}+Z'_\alpha g_{\mu\tau} J^\alpha_{\mu\tau}
	\end{eqnarray}
	Here, $D_\alpha=\partial_\alpha+iq_fg_{\mu\tau}Z'_\alpha$ is the covariant derivative with $q_f$ being the $U(1)_{L_\mu-L_\tau}$ charge for the respective particle. The $\mu-\tau$ current from eq. \eqref{zplag} is expressed as:
	\begin{eqnarray*}
		J_{\mu-\tau}^\alpha=q_f(\bar{\mu}\gamma^\alpha\mu+\bar{\nu_\mu}\gamma^\alpha P_L\nu_\mu-\bar{\tau}\gamma^\alpha\tau-\bar{\nu_\tau}\gamma^\alpha P_L\nu_\tau). 
	\end{eqnarray*}
	In eq. \eqref{zplag}, $Z'_{\alpha\beta }=(\partial_\alpha Z'_\beta-\partial_\beta Z'_\alpha)$ and $F_{\alpha\beta}$ are the field strength tensors of $U(1)_{L_\mu-L_\tau}$ and $U(1)_Y$ symmetry groups respectively with kinetic mixing parameter $\epsilon$. Even if this kinetic mixing is considered absent in the Lagrangian, it can arise at one loop level with particles charged under both the gauge sector in the loop and this mixing can be approximated to $\epsilon \simeq g_{\mu \tau}/70$ \cite{Escudero:2019gzq}. While the phenomenology of muon $(g-2)$, and DM relic in our model is not dependent on this mixing, the kinetic mixing will be tiny for our chosen parameter space. The VEVs of the singlet scalars are associated with the fate of the additional gauge boson that appears due to the $U(1)_{L_\mu-L\tau}$ symmetry, and the new gauge boson mass can be found to be $M_{Z'}=g_{\mu \tau} \sqrt{(1-n_X)^2v^2_1+v^2_2}$ with $g_{\mu \tau}$ being the $L_{\mu}-L_{\tau}$ gauge coupling.
	
	\subsection{ Neutrino mass}
	While there are several ways to realise sub-eV Dirac neutrino mass, we consider the simplest possibility of type-I seesaw for Dirac neutrinos in $L_{\mu}-L_{\tau}$ model. The additional particle content required to implement type-I Dirac seesaw is shown in table \ref{tab1}. The relevant Lagrangian is:
	\begin{align}
		-\mathcal{L}_{Y} \supset & Y_e \overline{L_e} \tilde{H} N_{eR} + Y_{\mu} \overline{L_{\mu}} \tilde{H} N_{\mu R} + Y_{\tau} \overline{L_{\tau}} \tilde{H} N_{\tau R} \nonumber \\
		& + M_1 \overline{N_{e L}} \nu_{eR}+ Y_{22} \overline{N_{\mu L }} \nu_{\mu R} \Phi_1 + Y_{33} \overline{N_{\tau L}} \nu_{\tau R} \Phi^{\dagger}_1 \nonumber \\
		& +Y_{12} \overline{N_{e L}} N_{\mu R} \Phi^{\dagger}_2 + Y_{13} \overline{N_{e L}} N_{\tau R} \Phi_2 + M_{R\alpha} \overline{N_{\alpha L}} N_{\alpha R}+ {\rm h.c.}
	\end{align}
	We also assume a global, unbroken lepton-number symmetry to ensure that Majorana mass terms for singlet fermions are absent\footnote{ Additionally, there can be direct coupling of $L_e$ and $\nu_{eR}$ via the SM Higgs, which we are ignoring. While this term contributes only to the overall neutrino mass, additional discrete symmetries can be introduced to forbid it.}.
	
	In $(\overline{\nu_{\alpha L}}, \overline{N_{\alpha L}})$ and $(\nu_{\alpha R},N_{\alpha R})^T$ basis (where $\alpha=e,\mu,\tau$), the $6\times6$ mass matrix for light neutrinos and heavy fermions can be written as,
	\begin{eqnarray}
		\begin{pmatrix}
			\overline{\nu_{\alpha L}}&\overline{N_{\alpha L}}\\
		\end{pmatrix}\begin{pmatrix}
			0&M_D'\\M_D&M_{LR}
		\end{pmatrix}\begin{pmatrix}
			\nu_{\alpha R}\\N_{\alpha R}
		\end{pmatrix},
	\end{eqnarray}
	
	where 
	\begin{align}
		&  M'_D=\frac{v_h}{\sqrt{2}}\begin{pmatrix} Y_e & 0 & 0\\
			0 & Y_{\mu} & 0 \\
			0 & 0 & Y_{\tau}\end{pmatrix},   M_D=\frac{v_1}{\sqrt{2}}\begin{pmatrix} \frac{\sqrt{2}M_1}{v_1} & 0 & 0\\
			0 & Y_{22} & 0 \\
			0 & 0 & Y_{33}\end{pmatrix}, \nonumber \\
		& M_{LR} = \begin{pmatrix} M_{Re} & Y_{12} \frac{v_2}{\sqrt{2}} & Y_{13} \frac{v_2}{\sqrt{2}}\\
			Y_{21} \frac{v_2}{\sqrt{2}} & M_{R\mu} &0 \\
			Y_{31} \frac{v_2}{\sqrt{2}} & 0 & M_{R\tau}\end{pmatrix}.
	\end{align}
	Here $v_h, v_{1,2}$ are the vacuum expectation values (VEV) of the SM Higgs and two singlet scalars, respectively. The $M_{LR}$ mass term 
	arise due to $\Phi_2$ being naturally larger, which allows us to have light neutrino mass within the allowed range. Therefore, working within the limit $M_D, M_D'<<M_{LR}$, the light neutrino mass matrix is given by 
	\begin{equation}
		m_{\nu}=-M'_D M^{-1}_{LR} M_D.
	\end{equation}

	Clearly, the model predicts a diagonal charged lepton mass matrix\footnote{Yukawa interaction terms with the SM Higgs $``\overline{L_\alpha} H\alpha_R"$ for $\alpha=e,\mu,\tau$ are allowed and will give rise to a diagonal charged lepton mass matrix.} $M_\ell$ and diagonal Dirac Yukawa of light neutrinos. Thus, the non-trivial neutrino mixing will arise from the structure of $M_{LR}$ matrix only, which is generated by the chosen scalar singlet fields. 
	In an obvious way, there will be mixing among the heavy and light states, which gives rise to the new right-handed states as,
	\begin{eqnarray}
		&&\nu_{iR}=\cos\theta_v\nu_{\alpha R}+\sin\theta_v N_{\alpha R}\\
		\text{and}&& N_{iR}=\cos\theta_v N_{\alpha R}-\sin\theta_v\nu_{\alpha R},
	\end{eqnarray}
	where, $i=1,2,3$,  $\alpha=e,\mu,\tau$ and $\theta_v\sim \tan^{-1}(M_D.M_{LR}^{-1})$, which will be heavily suppressed due to the large $M_{LR}$ mass and light $M_{D}$ mass. Therefore, we can safely use the notation $\nu_R$ instead of $\nu_{iR}$ in the following sections. The heavy fields associated with neutrino mass generation decay and end much earlier due to their large masses, so they wouldn't affect the latter part of the study.
	We  keep the notation $\nu_{e R}$ for the $1^{st}$ generation of right-handed neutrino and use $\nu_{\beta R}$ with $\beta=\mu,\tau$ for the rest which are having non-zero $U(1)_{L_\mu-L_\tau}$ charge.
	
	A dedicated numerical scan of the neutrino sector has not been performed in the present work. Since the neutrino mass matrix depends on several free parameters, viable regions consistent with neutrino oscillation data can be obtained through appropriate choices of the Yukawa couplings and singlet-scalar vacuum expectation values. In this study, we focus instead on the dark matter and cosmological implications of the model, adopting benchmark values for the singlet-scalar vacuum expectation values that yield the desired $Z'$ mass throughout the analysis.

	\subsection{ Anomalous Muon Magnetic Moment}
	The magnetic moment of muon is given by
	\begin{equation}\label{anomaly}
		\overrightarrow{\mu_\mu}= g_\mu \left (\frac{q}{2m} \right)
		\overrightarrow{S}\,,
	\end{equation}
	where, $g_\mu$ is the gyromagnetic ratio and its value is $2$ for a structureless, spin $\frac{1}{2}$ particle of mass $m$ and charge $q$. Any radiative correction, which couples the muon spin to the virtual fields, contributes to its magnetic moment and is given by
	\begin{equation}
		a_\mu=\frac{1}{2} ( g_\mu - 2)
	\end{equation}
	
	\begin{figure}[h]
		\includegraphics[scale=0.8]{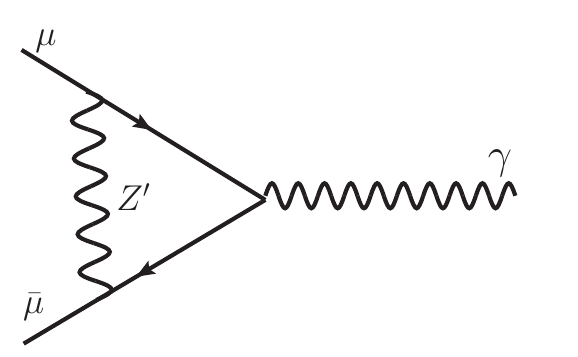}
		\caption{Contribution to muon anomalous magnetic moment via the new gauge boson $Z'$}\label{mmm}
	\end{figure}
	The anomalous muon magnetic moment has been measured very precisely, while it has also been predicted in the SM to great accuracy.
	In our model, the additional contribution to the muon magnetic moment comes from a one-loop diagram mediated by the $Z'$ boson, shown in fig. \ref{mmm}. The contribution is given by \cite{Brodsky:1967sr, Baek:2008nz, Queiroz:2014zfa}
	\begin{equation}
		\Delta a_{\mu} = \frac{\alpha'}{2\pi} \int^1_0 dx \frac{2m^2_{\mu} x^2 (1-x)}{x^2 m^2_{\mu}+(1-x)M^2_{Z'}} ,
	\end{equation}
	where $\alpha'=g^2_{\mu\tau}/(4\pi)$. \\
	\begin{figure}[h!]
		\includegraphics[scale=0.5]{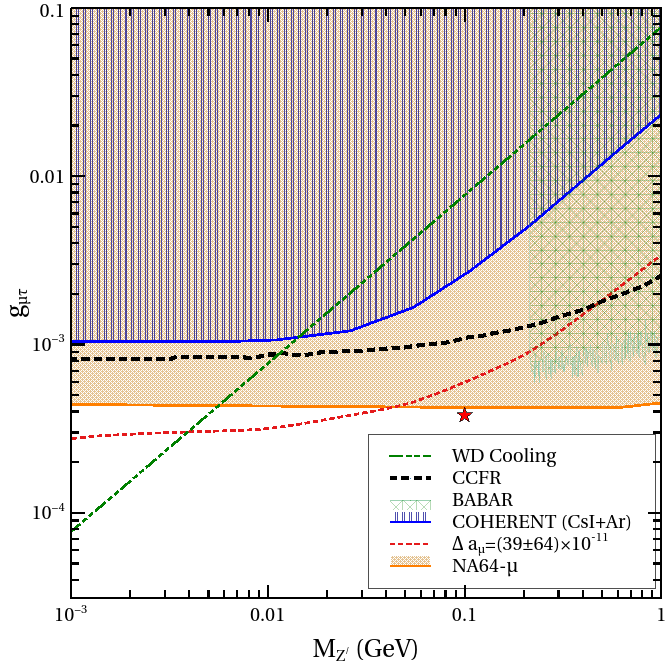}
		\caption{Summary plot containing related bounds as indicated in the figure key. The red star is the benchmark value that we have used for further analysis.}\label{bounds}
	\end{figure}
	
	The traditional discrepancy between the experimental value and the Standard Model prediction of the anomalous magnetic moment of the muon, $a_\mu$, has motivated a number of new-physics interpretations. Although recent lattice-QCD–based SM predictions \cite{Aliberti:2025beg} indicate a reduced tension, the Fermilab measurement \cite{Muong-2:2025xyk}, when combined with updated SM inputs, remains a key probe of possible physics beyond the Standard Model. The updated input at 1$\sigma$ deviation reads as $\Delta a_\mu^{\rm exp}=(39\pm64)\times 10^{-11}$.
	In figure \ref{bounds}, we have used various experimental bounds to choose a viable parameter space. The shaded grey-coloured exclusion band corresponds to the upper bound on the neutrino trident process measured by the CCFR
	collaboration \cite{Altmannshofer:2014pba}. The exclusion region labelled BABAR corresponds to the limits imposed by the BABAR collaboration \cite{BaBar:2016sci} on four-muon final states at high $Z'$ mass, shaded in light green. The astrophysical bounds from the cooling of the white dwarf (WD) \cite{Bauer:2018onh, Kamada:2018zxi} exclude the upper left triangular region. The observation of coherent elastic neutrino-nucleus cross section in combined liquid argon (LAr) and caesium-iodide (CsI) performed by the COHERENT Collaboration \cite{Akimov:2020pdx} is shown in a blue solid line and shades the excluded region above the blue line. The upper bound on $(g-2)_\mu$ corresponds to $\Delta a_\mu^{\rm exp}=(39\pm64)\times 10^{-11}$ is shown in a red dashed line. The latest bound from NA64-$\mu$ \cite{NA64:2024klw} is shown in the orange line, and the upper shaded region is the excluded region. For our further analysis in this work, we have chosen the undisturbed region, below the upper bounds, where $(g-2)_\mu$ results are consistent. Therefore, we focus on a benchmark region of the parameter space characterized by a gauge coupling of $g_{\mu\tau}=3.8\times10^{-4}$ and mediator masses in the range $1{\rm MeV}\leq M_{Z'}\leq 200{\rm MeV}$, consistent with current experimental constraints. For the dark matter study, we fix $M_{Z'}=100{\rm MeV}$ and $g_{\mu\tau}=3.8\times10^{-4}$ and divide the analysis into two distinct mass regimes discussed in respective sections.

	\section{Cosmological consequences}\label{sec3}
	
	\subsection{Fate of $\nu_R$}
\begin{figure}[h]
		\centering
		\includegraphics[scale=0.45]{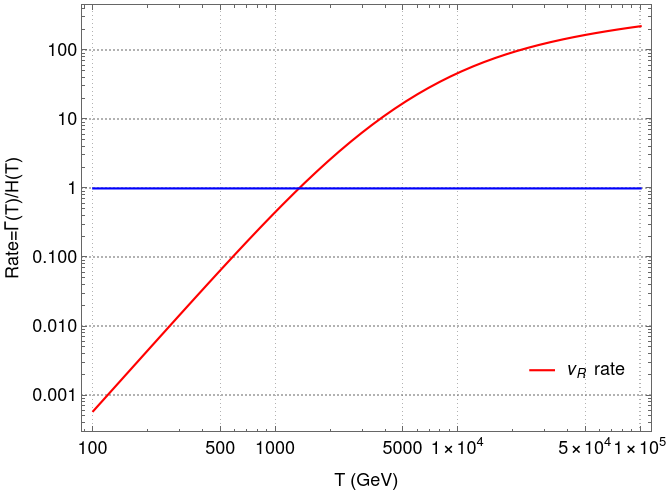}
		\caption{Interaction rate $vs.$ temperature for the light $\nu_{eR}$. The horizontal line indicates $\Gamma/H=1$.}
		\label{rate123}
	\end{figure}

	We have considered that $\nu_{eR}$ maintains thermal equilibrium with the heavy fields and SM contents via Yukawa interactions. The heavy fields ($N_R,\Phi_1,\Phi_2$) along with the lighter field ($\nu_{eR}$) leave the thermal equilibrium at a very high temperature and freeze out. The $t-$channel processes, such that $\nu_{eR}\nu_{eR}\leftrightarrow N_LN_L$ via $\Phi_2$ and $\nu_{eR}\nu_{eR}\leftrightarrow \Phi_2\Phi_2$ via $N_L$ processes will ensure the decoupling of the fields by comparing the interaction rate ($\langle\Gamma_{xx\rightarrow yy}\rangle\times n^{eq}_{target}$) with the Hubble rate. The Hubble rate varies as $H\propto \frac{T^2}{M_{\rm Planck}}$ and $n^{eq}_{target}$ is the equilibrium number density of the target particle. We have considered $M_{N}\sim\mathcal{O}(10^4)$ GeV, $M_{\Phi_1}\sim \mathcal{O}(10^2)$ GeV and $Y_{22}=Y_{33}\sim\mathcal{O}(10^{-3}$)\footnote{These choices of numerical values are consistent with neutrino oscillation data.} to check the decoupling profile of $N_R,\Phi_1$ and $\nu_R$, which are shown in Fig. \ref{rate123}. We can see that the light $\nu_{eR}$ decouples much earlier than the electroweak (EW) scale. Since $\nu_{eR}$ are very light, they will contribute to radiation density, hence there will be a non-zero contribution to ${\Delta N_{\rm eff}}$. Dirac neutrinos decoupling above the EW scale give a fixed contribution of $\Delta N_{\rm eff}=0.0466(=0.14/3)$ \cite{Luo:2020sho}, which is allowed from the current Planck 2018 bound \cite{Planck:2018vyg}.
	
\begin{figure}[h]
		\includegraphics[width=0.45\textwidth]{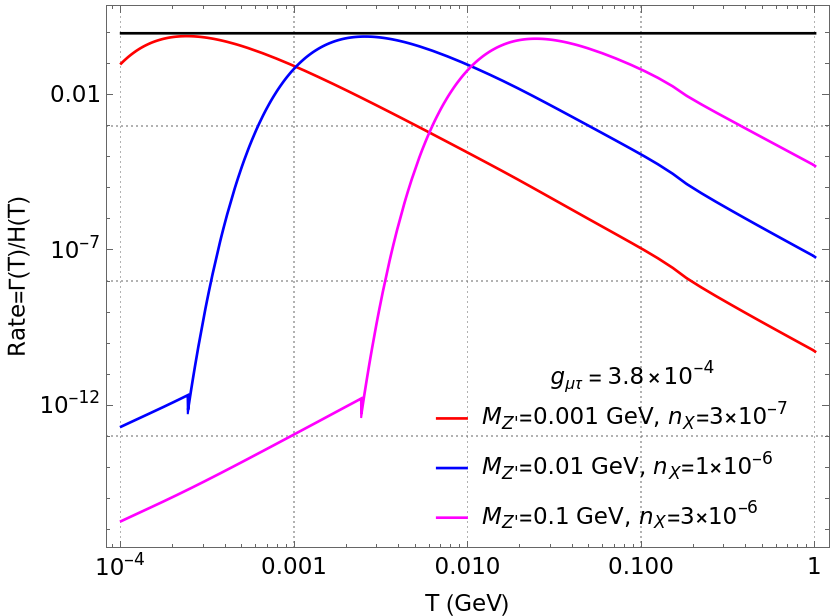}
		\caption{Interaction rate $vs.$ temperature for the new gauge boson mass $M_{Z'}$ and gauge charge, $n_X$ with fixed $g_{\mu\tau}$. The black horizontal line represents $\Gamma/H=1$. Hence, the region below this line indicates processes that never reach the thermal bath. }
		\label{nur1}
	\end{figure}

	As $\nu_{\beta R}$'s carry non-zero gauge charge, they may maintain thermal equilibrium with the SM plasma via gauge-mediated interactions as long as the interaction rate exceeds the Hubble expansion rate of our universe at a specific temperature. Interestingly, our choice of $Z'$ mass will spoil the whole study if $\nu_{\beta R}$ enters thermal equilibrium with the $Z'$ and decouples much later. The $\nu_{\beta R}$ enters the thermal equilibrium at $T\sim M_{Z'}$ and decouples later around $T\sim M_{Z'}/10$ \cite{Heeck:2012bz}. Decoupling of light $\nu_R$ below 200 MeV will increase the ${\Delta N_{\rm eff}}$ value significantly, which is strictly ruled out by Planck 2018 data \cite{Planck:2018vyg}. The $Z'$ mediated $\nu_R\overline{\nu_R}\leftrightarrow f\bar{f}$ interaction cross-section is evaluated as:
	\begin{eqnarray}
		\sigma(\nu_R\overline{\nu_R}\rightarrow f\bar{f})=\frac{k_f n_X^2q_f^2g_{\mu\tau}^4\sqrt{1-\frac{4M_f^2}{s}}(2M_f^2+s)}{12\pi(M_{Z'}^2-s)^2}.\label{cr1}
	\end{eqnarray}
	Here $k_f$ is the phase space factor, $n_X(q_f)$ is the new gauge charge for $\nu_{\beta R}(l_{\alpha})$.
	We can see from Eq. \eqref{cr1}, the cross-section is proportional to the gauge coupling, gauge charge and $Z'$ mass. We have considered the parameter space as $ 1 {\rm MeV}\le M_{Z'}\le 200$ MeV and the gauge coupling $g_{\mu\tau}= 3.8\times 10^{-4}$. To make sure that $\nu_{\beta R}$ does not enter the thermal equilibrium with $Z'$ and spoil the study, we can tune the gauge charge $n_X$ in such a way that $\nu_{\beta R}$ never reaches thermal equilibrium with the SM bath. In Fig. \ref{nur1}, we show the rate (interaction rate divided by Hubble rate) $vs.$ temperature for three benchmark values of $Z'$ mass and gauge charge $n_X$ with fixed gauge coupling $g_{\mu\tau}$. 
	For a gauge charge, $n_X<\mathcal{O}(10^{-6})$, the light $\nu_{\beta R}$ never reaches thermal equilibrium for our choice of parameter space and never contributes to $ \Delta N_{\rm eff}$ via the $Z'$ mediated $s$-channel process. However, such light $Z'$ can also decay to active neutrinos and contribute independently to $N_{\rm eff}$ value. We will discuss it in a later section.

	\subsection{Dark Matter and Relic Abundance}
	For the dark matter sector, we introduce one additional vector-like fermion $\psi$ with $L_{\mu}-L_{\tau}$ gauge coupling being $g_\psi$. The relevant Lagrangian can be written as follows:
	\begin{align}
		\mathcal{L} & \supseteq \overline{\psi} i \gamma^\mu D_\mu \psi - M_\psi \overline{\psi} \psi
	\end{align}
	Here $D_\mu \psi = (\partial_{\mu}+ig_\psi Z'_{\mu}) \psi$ and $g_\psi = n_\psi g_{\mu \tau}$ with $n_\psi$ being gauge charge of a vector-like fermion $\psi$. Since $n_\psi$ can be chosen independently, we keep $g_\psi$ as a free parameter. We have set a benchmark point for $Z'$ mass and coupling, consistent with $(g-2)_\mu$ results with $M_{Z'}=100$ MeV and $g_{\mu\tau}=3.8\times 10^{-4}$ to divide the parameter spacer into two distinct regions.

	For $M_{DM}> M_{Z'}$, the relevant annihilation cross section scales as the combination $g_{\mu\tau}^4/M_{DM}^2$ (for a fixed choice of gauge charges). Hence, the correct relic density depends on the gauge coupling and DM mass. On the other hand, for $M_{DM}<M_{Z'}$ the cross-section scales as the combination $g_{\mu\tau}^4M_{DM}^2/M_{Z'}^4$, and thus the right relic density selects a value for the ratio $g_{\mu\tau}/M_{Z'}$, for a given value of DM mass. In the second case, we particularly focus on the resonant region, where $M_{DM}\sim M_{Z'}/2$ to obtain the correct relic density. They are discussed in detail below.  
	
	\begin{itemize}
		\item  {\bf When $M_{DM}>M_{Z'}$ :}
		
		\begin{figure}[h]
			\includegraphics[scale=0.6]{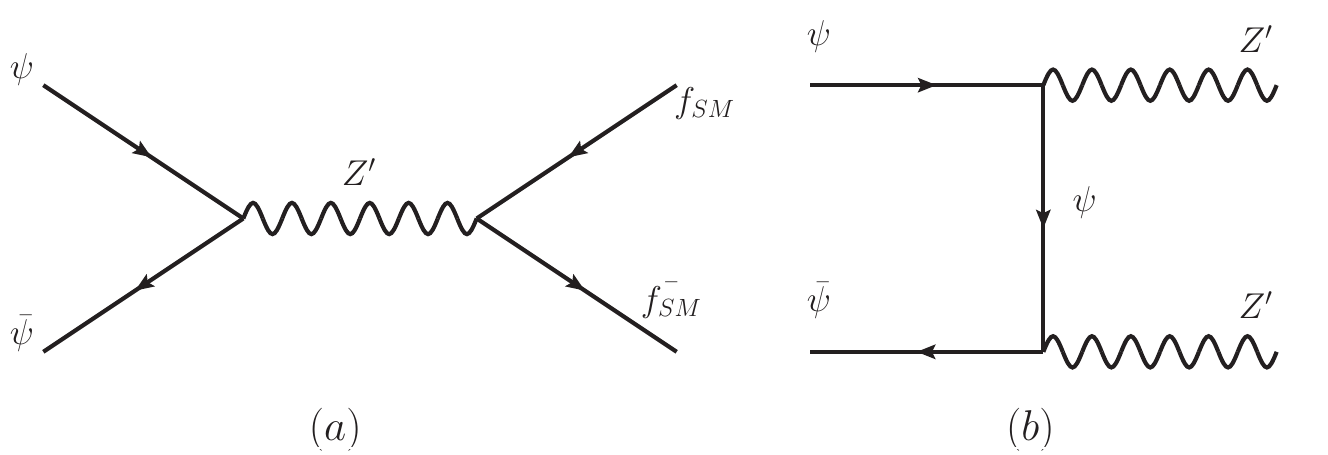}
			\caption{Dominant contribution to the dark matter annihilation processes via $Z'$}
		\end{figure}
		The $s-$channel contribution to the DM annihilations are given by \cite{Altmannshofer:2016jzy}:
		\begin{eqnarray}
			\langle\sigma v\rangle({\psi\bar{\psi}\rightarrow f\bar{f}})= q_f^2\frac{n_\psi^2 g_{\mu\tau}^4}{2\pi}\sqrt{1-\frac{m_f^2}{M_\psi^2}}\frac{2M_\psi^2+m_f^2}{(4M_\psi^2-M_{Z'}^2)^2}\\
			\langle\sigma v\rangle({\psi\bar{\psi}\rightarrow \nu\bar{\nu}})= q_f^2\frac{n_\psi^2 g_{\mu\tau}^4}{2\pi}\frac{M_\psi^2}{(4M_\psi^2-M_{Z'}^2)^2}
		\end{eqnarray}
		At the same time, the $t-$ channel contribution is given by \cite{Altmannshofer:2016jzy}:
		\begin{eqnarray}
			\langle\sigma v\rangle({\psi\bar{\psi}\rightarrow Z'Z'})=\frac{g_\psi^4}{16\pi M_\psi^2}\Big(1-\frac{M_{Z'}^2}{M_\psi^2}\Big)^{3/2}\Big(1-\frac{M_{Z'}^2}{2M_\psi^2}\Big)^{-2}
		\end{eqnarray}
		In the above expressions of thermal averaged cross-section, $q_f$ stands for the lepton charge under the $U(1)_{L_\mu-L_\tau}$ symmetry and $g_\psi=n_\psi g_{\mu\tau}$. The dominant contribution to the DM relic is coming from the $s-$channel processes. Finally, the total contribution to the thermally averaged cross-section is given by the sum of all three contributions as:
		\begin{eqnarray}
			\langle\sigma v\rangle=&&\langle\sigma v\rangle({\psi\bar{\psi}\rightarrow f\bar{f}})+\langle\sigma v\rangle({\psi\bar{\psi}\rightarrow \nu\bar{\nu}})+\langle\sigma v\rangle({\psi\bar{\psi}\rightarrow Z'Z'})
		\end{eqnarray}
		\begin{figure}
			\includegraphics[scale=0.5]{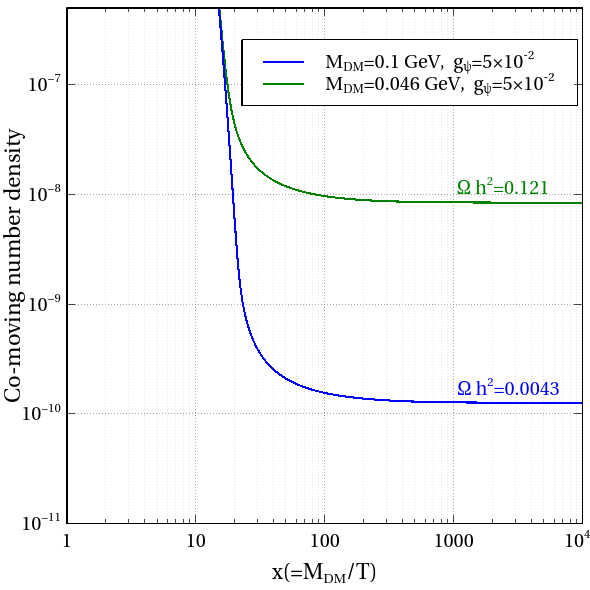}
			\caption{Comoving number density for the DM candidate with two benchmark parameters.}\label{comoving}
		\end{figure}
		The Boltzmann equation for the DM candidate for $n_{DM}=n_{\overline{DM}}$ with $n=n_{DM}+n_{\overline{DM}}$ can be expressed as:
		\begin{eqnarray}
			\frac{dn}{dt}+3nH=-\frac{1}{2}\langle\sigma v\rangle(n^2-n^2_{eq}).
			\label{be1}	\end{eqnarray}
		In the thermal freeze-out case, the DM particles were initially maintaining thermal equilibrium with the SM fluid via $2\leftrightarrow2$ scattering processes and after the freeze-out temperature $T_F$, they can no longer annihilate and the amount of relic abundance is given by:
		\begin{eqnarray}
			\Omega h^2=\frac{1.04 \times10^9 x_F}{\sqrt{g_*}M_{\rm Pl}\langle\sigma v\rangle}\text{GeV}^{-1},
		\end{eqnarray} 
		with $x_F=\frac{M_\psi}{T_F}~, M_{\rm Pl}=1.22\times 10^{18}$ GeV being the reduced Planck mass, $g_*$ being the relativistic degree of freedom, and $H$ being the Hubble parameter.
		
		The comoving number density for the DM is shown for various input parameters in Fig. \ref{comoving}. The green line corresponds to $M_{DM}=0.046$ GeV, which is close to the resonance region, triggering a sudden rise in the comoving number density (compared to the other two benchmark values) and giving rise to a relic abundance in the exact ballpark.

		\item {\bf When $M_{DM}<M_{Z'}$:}
		
		We now check the region where dark matter mass is less than the $Z'$ mass. 
		
		When DM is lighter than the $Z'$ boson, then the $Z'\rightarrow \psi\bar{\psi}$ decay channels will contribute to the total annihilation cross-section. 
		The partial decay width $Z'\rightarrow f\bar{f}$ is given by:
		\begin{eqnarray*}
			\Gamma_{Z'\rightarrow f\bar{f}}=k_f\frac{q_f^2g_{\mu\tau}^2M_{Z'}}{12\pi}\Big(1+2\frac{M_f^2}{M_{Z'}^2}\Big)\sqrt{1-\frac{4M_f^2}{M_{Z'}^2}},
		\end{eqnarray*}
		with $k_f=1/2$ for neutrinos, else $k_f=1$ for other fermions\footnote{$k_f=\frac{1}{n!}$ is the phase space factor, which is multiplied to the decay width where there are $n$ identical particles in the final state. }, $q_f$ is the $U(1)_{L_\mu-L_\tau}$ charge for the respective fermions. Now, the $Z'$ mediated $s-$channel cross-section for the $\psi\bar{\psi}\rightarrow f\bar{f}$ processes is given by \cite{Holst:2021lzm}:
		\begin{eqnarray}
			\sigma(s)=\sigma_f\frac{n_{\psi}^2q_f^2k_fg_{\mu\tau}^4\beta_f}{12\pi s\beta_\psi}\Big[\frac{(s+2M_\psi^2)(s+2M_f^2)}{(s-M_{Z'}^2)^2+M_{Z'}^2\Gamma_{Z'}^2}\Big],
		\end{eqnarray}
		where $s$ is the Mandelstam variable, $\beta_i=\sqrt{1-4M_i^2/s}$ and $\Gamma_{Z'}=\Gamma_{Z'\rightarrow f\bar{f}}+\Gamma_{Z'\rightarrow \psi\bar{\psi}}$ with $f=\mu,\tau,\nu_{\mu,\tau}$\footnote{In this case, $Z'$ decay to only neutrinos will be possible due to the choice of $Z'$ mass ($\sim0.1$ GeV).}. For our convenience, we have defined a parameter $r$ as the ratio of the $Z'$ mass to the dark matter mass, $i.e.,$ $r=M_{Z'}/M_{\psi}$. The thermal-averaged cross-section has been evaluated following the approach of \cite{Gondolo:1990dk} and by solving the same Boltzmann equation from eq. \eqref{be1}, we numerically determine the DM relic as a function of $r(=M_{Z'}/M_{\psi})$. 
		\begin{figure}
			\includegraphics[scale=0.5]{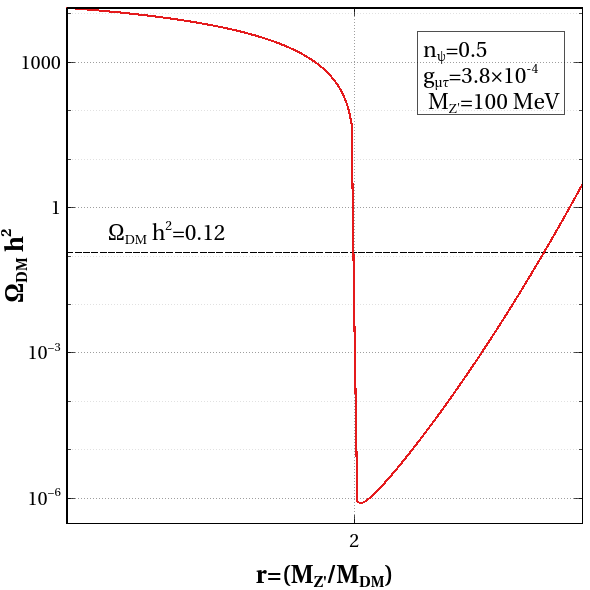}
			\includegraphics[scale=0.5]{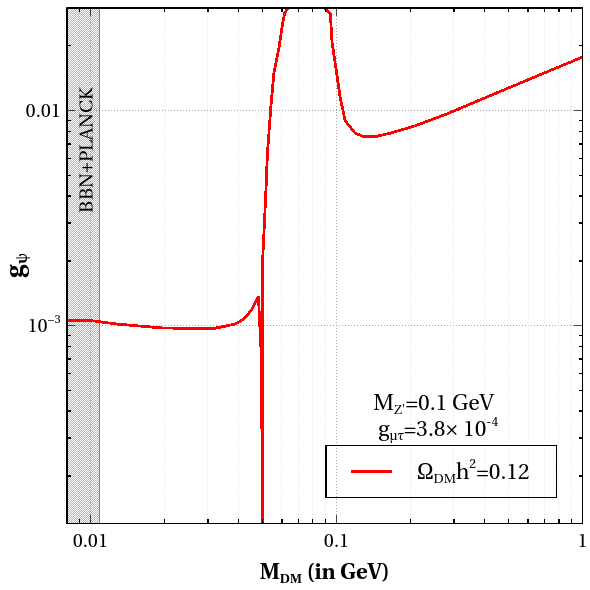}
			\caption{$(Left)$: Variation of DM relic with the mass ratio for a chosen benchmark point. Due to the resonance effect, there are two intersecting points in the dashed horizontal line. $(Right)$: A red solid contour curve satisfying $\Omega_{{\rm DM}} h^2=0.12$ in the gauge coupling ($g_\psi$) $vs.$ dark matter mass ($M_{DM}$) plane for the benchmark point. The vertical shaded region is from the BBN+PLANCK bound on the Dirac DM candidate \cite{Sabti:2019mhn}.}\label{resonant}
		\end{figure}
		
		In the first figure of Fig. \ref{resonant}, we show the dark matter relic abundance as a function of the mass ratio $r (=\frac{M_{Z'}}{M_\psi})$. We fixed the $Z'$ boson mass, coupling and gauge charge here to evaluate the relic evaluation pattern.  The black dotted line gives $\Omega_{DM}h^2=0.12$, and the red contour line shows the variation of DM relic abundance as a function of $r$. We can find that the red contour line crosses the dotted line on the DM relic abundance at two points. At $r\sim2$, the DM annihilation cross-section increases abruptly due to the Breit-Wigner enhancement \cite{Ibe:2008ye}, resulting in a sudden dip in the relic abundance pattern. After hitting the resonance dip, it increases slowly due to active decay channels present in the system and crosses the horizontal line at $r\sim2.71$. 
		
		In the second Fig. \ref{resonant}, we have shown a contour line and a region satisfying the current best-fit value for DM $\Omega h^2=0.12$ in the gauge coupling ($g_\psi$) vs. DM mass ($M_{DM}$) plane for the chosen benchmark value and the region from the summary plot. A discontinuity in the pattern is observed exactly near $M_{DM}=50$ MeV, due to resonance and the kink around $M_{DM}\sim100$ MeV appears as muons in the final state start contributing to the DM annihilation processes. 
		
		It is to be noticed that when such a light DM candidate is in thermal equilibrium during BBN, it can annihilate into the electron-positron pair, giving rise to an unacceptably large contribution to the expansion rate of the Universe, which severely affects the primordial element abundances measured
		today. Apart from this, CMB measurements also set a tight constraint on light DM entropy transferred into electrons after neutrinos decouple. The combined results from BBN+PLANCK put a lower bound on light Dirac DM mass up to $M_{DM}>10.9$ MeV \cite{Sabti:2019mhn}, which is shaded with the vertical light-red coloured band.
		
		If DM mass is larger than the muon mass (in the earlier case, when $M_{\rm DM}>M_{Z'}$), they can dominantly annihilate into muons; however, in such a case, there is a lower bound on sub-GeV DM, which is completely ruled out for $M_{DM}<M_\mu$. 
		We note that for dark matter masses above the dimuon threshold, the annihilation channel $\psi\bar{\psi}\rightarrow\mu^+\mu^-$ becomes kinematically accessible. Since the annihilation process is predominantly s-wave, energy injection during the recombination epoch may lead to additional constraints from CMB anisotropy measurements. A dedicated analysis of these bounds lies beyond the scope of the present work and may further restrict portions of the higher-mass parameter space. However, the low-mass resonant region, which constitutes the primary focus of our cosmological analysis, is largely unaffected since annihilation proceeds predominantly into neutrinos below the dimuon threshold.
	\end{itemize}
	
	\subsubsection{Direct Detection of DM:}
	\begin{figure}[h]
		\includegraphics[scale=0.12]{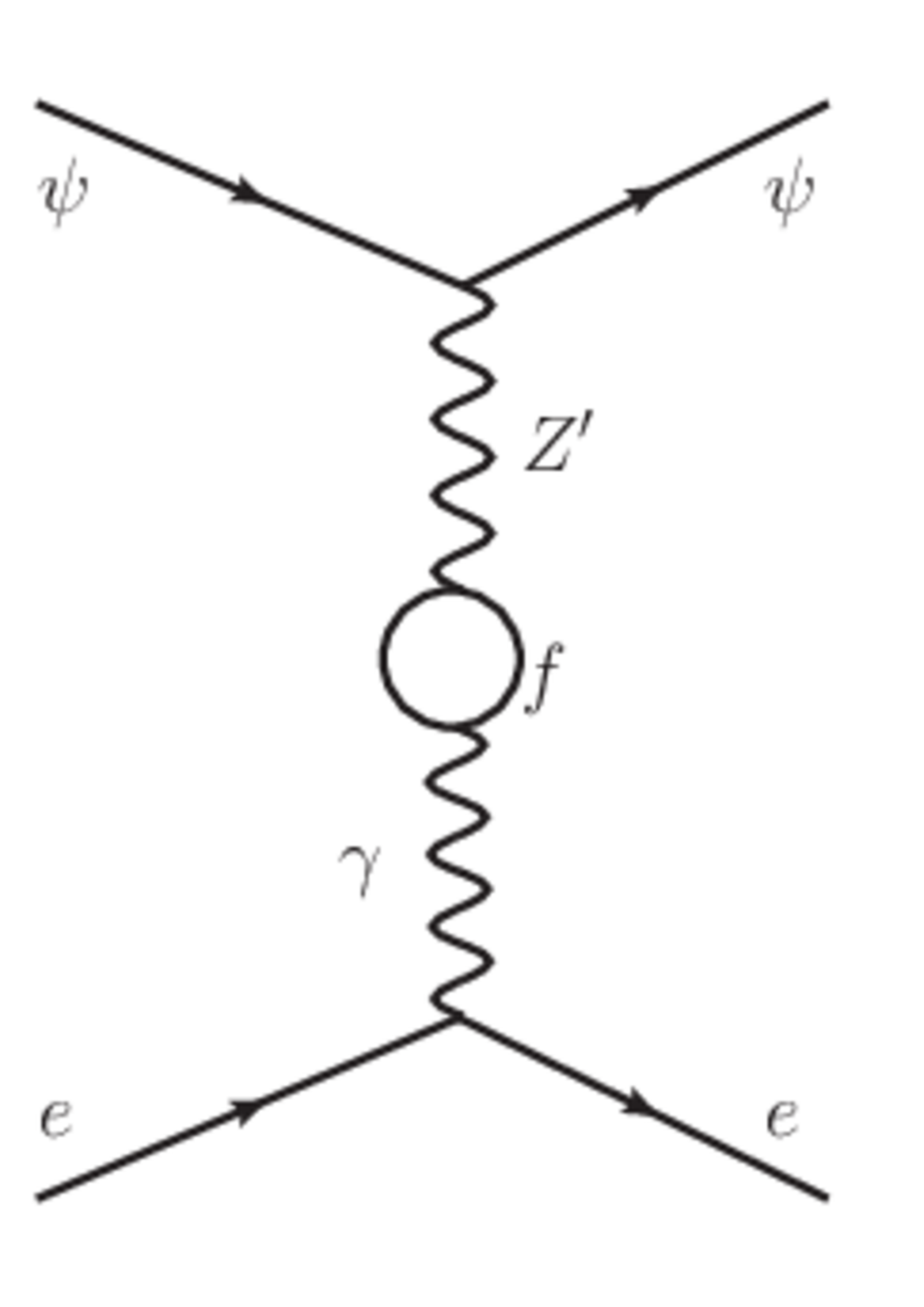}
		\includegraphics[scale=0.5]{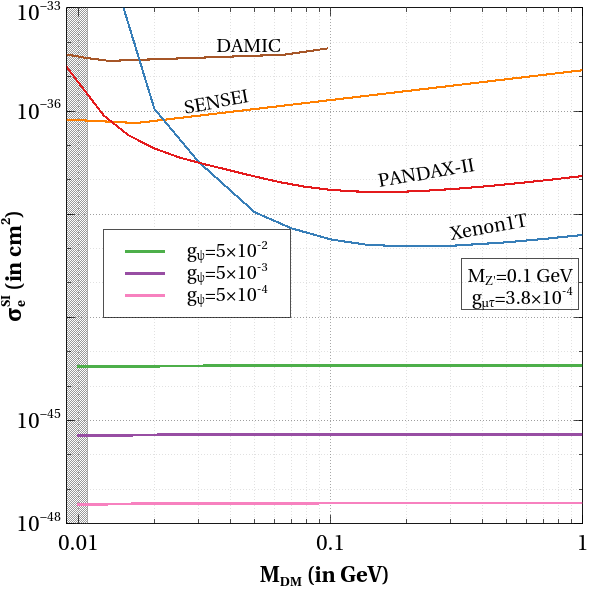}
		\caption{($Left$)DM-electron elastic scattering via one looped kinetic mixing between $Z'$ and the photon via $f=\mu,\tau$. ($Right$) Variation of spin-independent elastic scattering cross-section for $\psi e\rightarrow\psi e$ process with DM mass. The vertical grey shaded region is the BBN+PLANCK bound on the Dirac DM candidate, restricting $M_{DM}<10.9$ MeV \cite{Sabti:2019mhn}. We  show existing experimental constraint on spin-independent $DM-e$ scattering cross-sections from DAMIC \cite{DAMIC:2019dcn}, SENSEI \cite{SENSEI:2020dpa}, PANDAX-II \cite{PandaX-II:2021nsg} and XENON1T \cite{XENON:2019gfn}}\label{dd1}
	\end{figure}
	
	In the chosen mass range of interest throughout this study, the DM candidate does not couple directly to quarks but can have kinetic mixing between $Z'$ and the photons via $U(1)_{L_\mu-L_\tau}$ charged fermions as shown in the left panel of Fig. \ref{dd1}.
	
	Therefore, the DM-electron elastic scattering cross-section is given by \cite{Drees:2021rsg,Essig:2015cda}:
	\begin{eqnarray}
		\sigma(\psi e \rightarrow\psi e)=\frac{\mu_e^2}{\pi}\Big[\frac{g_\psi\epsilon e}{(M_{Z'}^2+\alpha_{EM}m_e^2)}\Big]^2;
	\end{eqnarray}
	where $\mu_e$ is the $\psi-e$ reduced mass and $\epsilon\simeq -\frac{g_{\mu\tau}}{70}$ is the strength of the kinetic mixing.
	
	We show the variation of the spin-independent cross-section with DM mass from Fig. \ref{dd1} for three benchmark $g_\psi$ values. This obvious suppression of cross-section values is due to the factor $m_e/M_{Z'}<<1$. 
	
	\subsection{Resonance and early kinetic decoupling}
	In conventional WIMP scenarios, the kinetic decoupling temperature is typically much smaller than the freeze-out temperature ($T_{\rm KD}\ll T_{\rm FO}$). However, in the present framework the observed relic abundance is achieved through Breit-Wigner resonance enhancement, corresponding to $M_\psi \simeq M_{Z'}/2$, where the annihilation cross section is significantly enhanced only in the vicinity of $s\simeq M_{Z'}^2$ \cite{Heeck:2014zfa}.
	
	In such resonant scenarios, the annihilation process can remain efficient while elastic scattering does not receive the same enhancement. Consequently, the momentum-transfer rate may fall below the Hubble expansion rate at an earlier epoch, leading to kinetic decoupling significantly earlier than in the standard WIMP case \cite{Binder:2017rgn}. To examine this possibility, we compute the scattering rate for the process $\psi\nu \rightarrow \psi\nu$ mediated by $Z'$ and compare it with the Hubble expansion rate. The approximate ratio is given by
	\begin{equation}
		\frac{\Gamma}{H}\simeq 0.021,n_\psi^2 g_{\mu\tau}^4
		\frac{M_{\rm Pl}T^4}{\sqrt{g_*}M_{\rm DM}M_{Z'}^4}.
	\end{equation}
	
\begin{figure}[h]
		\centering
		\includegraphics[width=0.45\linewidth]{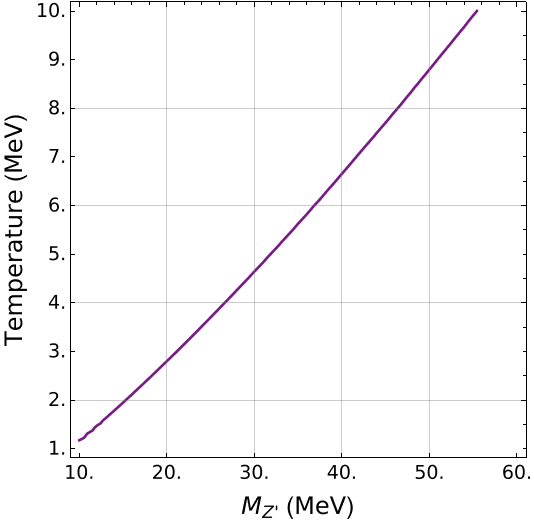}
		\includegraphics[width=0.45\linewidth]{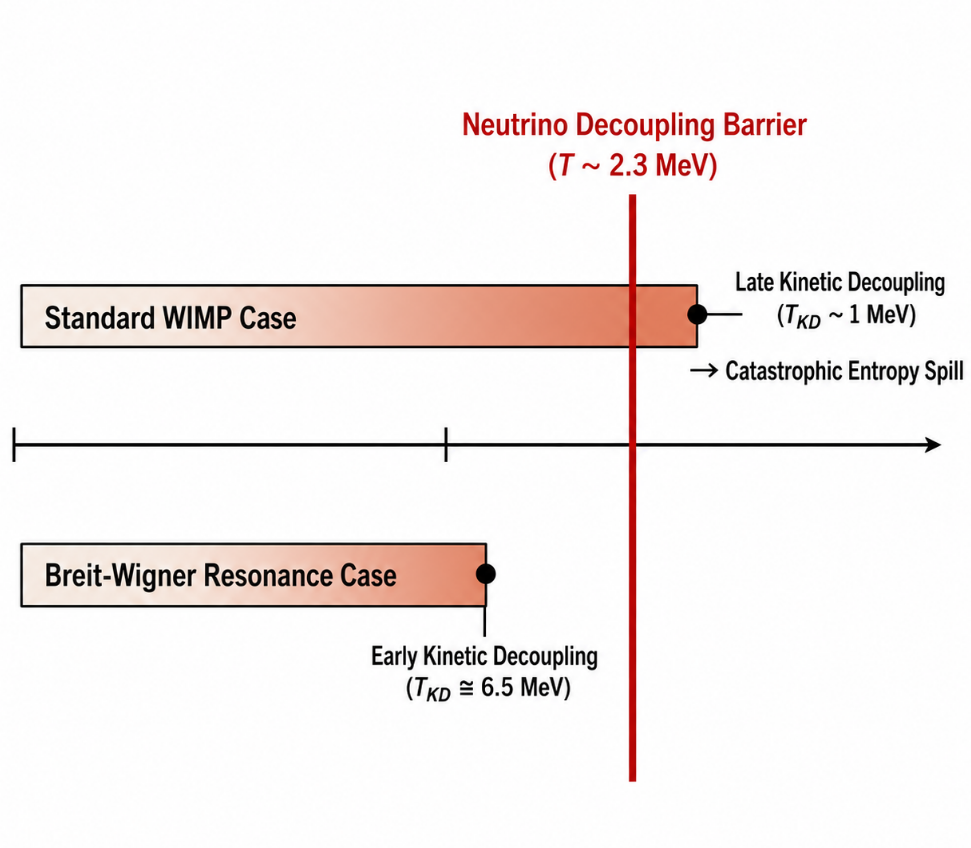}
		\caption{($Left$) The contour line representing $\Gamma/H=1$ in the given plane, where $\Gamma$ is the scattering rate for $\psi\nu\rightarrow \psi{\nu}$ via $Z'$ and $H$ being the Hubble rate. The region above the contour ensures the kinetically coupled region, while the region below is the decoupled region. ($Right$) A graphical representation showing for a benchmark value $M_{Z'}=40$ MeV ($\sim M_{\psi}=20$ MeV), the resonance enhancement triggering early kinetic decoupling. }
		\label{kdc}
	\end{figure}

	Using our benchmark parameter values, the resulting decoupling profile in the $M_{Z'}$--$T$ plane is shown in the left panel of Fig.~\ref{kdc}, where the contour corresponds to $\Gamma/H=1$. The region above the contour represents $\Gamma/H>1$, where dark matter remains kinetically coupled to the neutrino bath, while the region below the contour corresponds to $\Gamma/H<1$, indicating kinetic decoupling. 
	As shown in the figure, kinetic decoupling indeed occurs considerably earlier than in the conventional freeze-out picture. For example, for $M_\psi\simeq20~{\rm MeV}$ (corresponding to $M_{Z'}\simeq40~{\rm MeV}$), kinetic decoupling takes place at $T_{\rm KD}\simeq6.5~{\rm MeV}$, whereas in a typical WIMP scenario one would expect $T_{\rm KD}\sim M_\psi/20\simeq1~{\rm MeV}$. This benchmark point is illustrated in the right panel of Fig. \ref{kdc} for better understanding.
	
	The early kinetic decoupling induced by the Breit--Wigner resonance enhancement may, in principle, modify the standard relic-density calculation based on conventional chemical freeze-out. After kinetic decoupling, the DM is no longer in kinetic equilibrium with the thermal bath and instead develops a separate temperature evolution, characterized by $T_\psi$. To consistently assess the impact of this effect, one needs to solve the coupled Boltzmann equations governing the DM number density and temperature evolution.
	
	Nevertheless, for the benchmark point considered here, the coupled evolution shows that the resulting modification of the thermally averaged annihilation rate does not appreciably alter the integrated annihilation history. In Appendix~\ref{apendix1}, we explicitly present the coupled Boltzmann treatment and demonstrate that the corresponding correction to the relic abundance is found to be approximately 0.22\%, and is therefore negligible for the parameter space considered. We therefore retain the DM relic-density analysis presented in the earlier section unchanged.
	
	\subsection{Contribution to $\Delta N_{\rm eff}$ from the decay of light species:} 
	For the situation where $M_{DM}<M_{Z'}$, (as we set the benchmark value of $M_{Z'}=100$ MeV), the DM and the $Z'$ can transfer their entropy to the neutrinos after decoupling from the bath. Interestingly, if they (DM $\&~Z'$) can dump their entropy after the neutrinos decouple from the photons ($T_{\nu,D}\sim 2.3$ MeV), they can influence the Big Bang Nucleosynthesis (BBN) by triggering the Hubble expansion rate of the Universe. To justify the separate treatment of the $Z'$ and DM contributions to $N_{\rm eff}$, it is important to verify that the DM remains in kinetic equilibrium only until sufficiently early times and undergoes kinetic decoupling before neutrino decoupling. As discussed in the previous section, $T_{KD}$ lies well above the neutrino decoupling temperature, $T_{\nu, D}\simeq2.3$ MeV, indicating that the DM sector is already kinetically decoupled when neutrinos decouple. Consequently, the subsequent DM temperature evolution is no longer tied to the neutrino temperature, and the effects of early kinetic decoupling should be incorporated in the entropy-transfer calculation. While the entropy-conservation framework itself remains valid, the modified DM thermodynamic evolution can lead to a correction to the resulting $N_{\rm eff}$ calculated by the standard entropy transfer approach. In the following analysis, we will see that that this correction is negligible for larger $M_{Z'}$, where the standard and early-kinetic-decoupling treatments give nearly identical results, but becomes increasingly relevant for smaller $M_{Z'}$.
	
	In this part of the work, we will primarily focus on the impact of the new $Z'$ boson and the DM ($\psi$) on the neutrino energy density, which is generally parameterized by $N_{\rm eff}$.  The contribution from dark matter annihilation and $Z'$ decay to ${\rm N_{eff}}$ via radiation density is given by \cite{Drees:2021rsg}:
	\begin{eqnarray}
		N_{\rm eff}=N_\nu\Big[1+\frac{1}{N_\nu}\sum_{i=\psi,Z'}\frac{g_i}{2}I_k\big(\frac{M_i}{T_{\nu,D}}\big)\Big]^{4/3},\label{nefeq}
	\end{eqnarray}
	where, $g_i$ is the relativistic degrees of freedom for respective particles and the function $I_k(x)$ describe the entropy carried by the specific gauge boson ($k=g$) or fermion ($k=f$) at neutrino decoupling temperature, which is then transferred to the neutrinos. This is defined as \cite{Boehm:2013jpa}:
	\begin{eqnarray}
		I_k(x)=\frac{30}{7\pi^4}\int_{x}^{\infty}dy \frac{(4y^2-x^2)\sqrt{y^2-x^2}}{e^y\pm1};
	\end{eqnarray}
	the $+(-)$ sign refers to fermion (boson) statistics.

	However, when the DM kinetically decouples from the thermal bath before neutrino decoupling, its phase-space distribution no longer follows the equilibrium distribution characterized by the bath temperature. In this case, the DM temperature $T_\psi$ must be evolved independently together with the DM abundance. We therefore replace the equilibrium entropy contribution by the actual entropy density of the kinetically decoupled DM distribution evaluated at $T_{\nu,D}$.
	
	After kinetic decoupling, the DM develops a separate temperature $T_\psi$ and can be treated as a non-relativistic Maxwell--Boltzmann gas. Its entropy density is given by
	\begin{equation}
		s_\psi^{\rm KD}=\frac{\rho_\psi+P_\psi-\mu_\psi n_\psi}{T_\psi}.
	\end{equation}
	Using $\rho_\psi\simeq n_\psi m_{\rm DM}
	+\frac{3}{2}n_\psi T_\psi,$ and $
	P_\psi=n_\psi T_\psi$,
	together with the Maxwell--Boltzmann number density, one obtains
	\begin{equation}
		s_\psi^{\rm KD}=n_\psi
		\left[
		\frac{5}{2}
		+
		\ln\left(
		\frac{g_{\rm DM}}{n_\psi}
		\left(\frac{m_{\rm DM}T_\psi}{2\pi}\right)^{3/2}
		\right)
		\right].
	\end{equation}
	Here, the $5/2$ term arises from the non-relativistic kinetic-energy contribution ($3/2$) and the pressure contribution ($1$).

	We consequently redefine the fermionic entropy factor entering the $N_{\rm eff}$ calculation as
	\begin{eqnarray}
		I_f^{\rm KD}(m_{\rm DM})
		\equiv
		\frac{s_\psi^{\rm KD}(T_{\nu,D})}
		{s_{\nu}(T_{\nu,D})}    
	\end{eqnarray}
	with $s_{\nu}(T_{\nu,D})=
	\frac{7\pi^2}{90}T_{\nu,D}^3.$
	
	Notice that $I_f^{\rm KD}$ already contains the total number of DM internal degrees of freedom through $g_{\rm DM}=4$. Therefore, unlike the equilibrium expression, no additional factor of $g_{\rm DM}/2=4/2$ is required in the $N_{\rm eff}$ calculation given by Eq. \eqref{nefeq}.
	
	The resulting effective number of neutrino species after kinetic decoupling can therefore be reformed as, 
	\begin{eqnarray}
		N_{\rm eff}^{\rm KD}=
		N_\nu\left[
		1+\frac{1}{N_\nu}
		\left(
		I_f^{\rm KD}
		+\frac{3}{2}I_g(M_{Z'},T_{\nu,D})
		\right)
		\right]^{4/3}.   
	\end{eqnarray}
	
	This prescription consistently incorporates the depletion of the DM entropy after early kinetic decoupling. In particular, once the DM becomes sufficiently non-relativistic and its abundance is Boltzmann suppressed, $I_f^{\rm KD}$ becomes negligible. In this limit, $
	I_f^{\rm KD}\rightarrow0,$ and consequently $
	N_{\rm eff}^{\rm KD}
	\rightarrow
	N_{\rm eff}^{Z'},$ where $
	N_{\rm eff}^{Z'}=
	3\left[
	1+\frac{1}{3}
	\left(
	\frac{3}{2}I_g(M_{Z'},T_{\nu,D})
	\right)
	\right]^{4/3}.
	$
	
	Thus, we can expect overlapping  results between the kinetically decoupled DM prediction and the $Z'$ decay prediction in the parameter region, where $I_f^{\rm KD}\ll1$ is a direct consequence of the negligible residual DM entropy at neutrino decoupling, rather than an indication that the DM is absent from the preceding thermal evolution.
	
	\begin{figure} [h]
		\includegraphics[scale=0.534]{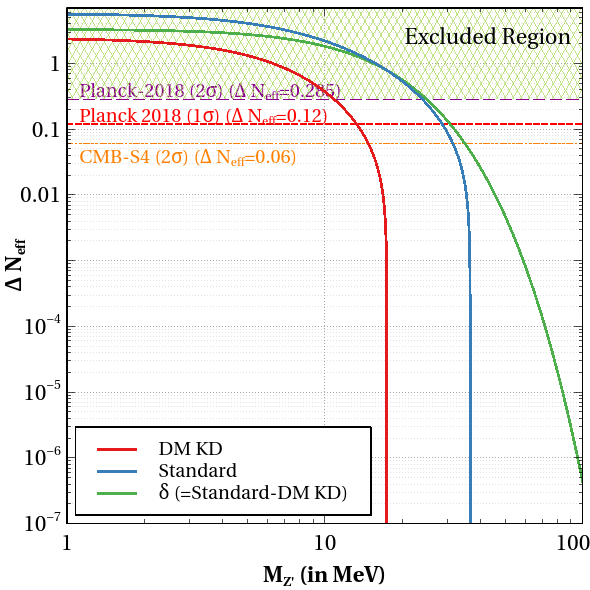}
		\includegraphics[scale=0.46]{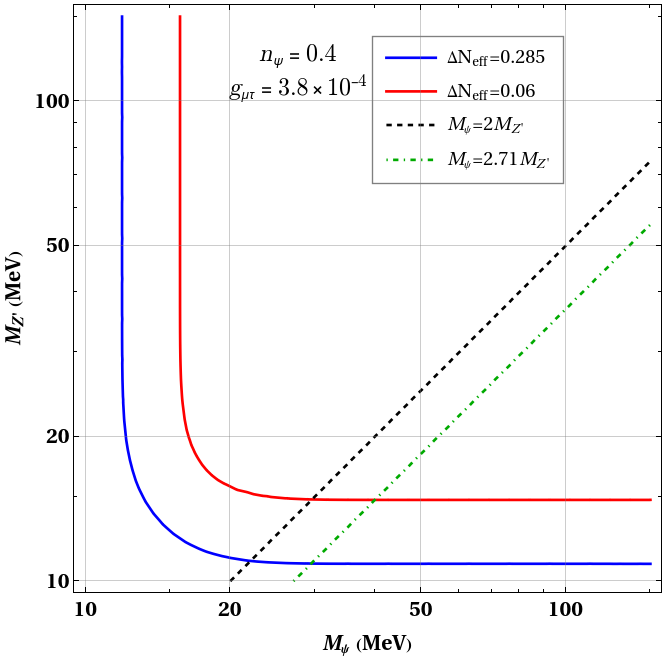}    
		\caption{($Left:$)Variation of $\Delta N_{\rm eff}$ with $Z'$ mass for different situations as indicated by the key text. The red one represents the contributions when the DM undergoes kinetic decoupling (KD) prior to the chemical decoupling due to the resonance effect. The blue curve represents the standard case if the resonance effects doesn't allow early kinetic decoupling. Finally, the green curve represents the difference between them. ($Right:$) Two contours corresponding to two $\Delta N_{\rm eff}$ values in the $M_{Z'} ~vs.~ M_\psi$ plane. The two dotted lines represents two resonance benchmark points satisfying DM relic. }\label{neff2}
	\end{figure}
	
	We used the general definition $ \Delta N_{\rm eff}=N_{\rm eff}-N_{\rm eff}^{\rm SM}$ to see the variation with our choice of parameters in this model. The current upper bound from Planck-2018 results restricts $ \Delta N_{\rm eff}\le0.285$. The combined results from SPT-3G and CMB-S4 restrict this value upto ${\Delta N_{\rm eff}}\le0.06$.

	To ensure the early KD contributions are sufficient to get the correct $N_{\rm eff}$ values, we have also checked the residual annihilation rate at neutrino
	decoupling. The ratio $R_{\rm ann}(T_{\nu, D})
	=\left.
	\frac{\Gamma_{\rm ann}}{H}
	\right|_{T=T_{\nu, D}}
	$ is found to be smaller than unity in the parameter region considered. Thus, annihilations are already out of chemical equilibrium at neutrino decoupling. Nevertheless, since $R_{\rm ann}<1$ does not by itself imply that the subsequent entropy transfer is completely negligible, we explicitly quantify the residual effect through $\delta$ (Standard- KD) in the figure.
	
	The left panel of Fig. \ref{neff2} shows variation of $\Delta N_{\rm eff}$ with the $Z'$ mass. Here the red curve represents the contributions to $\Delta N_{\rm eff}$ when DM undergoes early kinetic decoupling due to the resonance effect. Interestingly, if we plot the same curve for the $Z'\rightarrow \nu\nu$ decay, it will overlap with the red curve, as the integral function $I_f^{KD}\rightarrow0$. The blue curve represents the similar contributions but considers resonance effects not triggering early kinetic decoupling (KD), and their difference is shown by the green curve. 
	This figure shows that if only $Z'$ decay is producing neutrinos after they leave the thermal bath or KD is triggered by the resonance effects, then $M_{Z'}\le10.1$ MeV will give larger $\Delta N_{\rm eff}$ values which are ruled out by Planck 2018 data.  On the other hand, if light DM haven't left the thermal bath due to early KD in the same process, then this mass bound is further stretched towards the larger $M_{Z'}$ value. This enhancement to $\Delta N_{\rm eff}$ value is due to the presence of additional light DOFs, which will be there in the bath for a longer time and continue to contribute to the total energy density. We find that the correction becomes rapidly suppressed toward larger $M_{Z'}$, whereas a non-negligible deviation can occur in the lower-$M_{Z'}$ region. Therefore, for the region where the correction is small, the standard equilibrium entropy-transfer expression provides an excellent approximation. However, at the lower end of the resonant region, we
	instead use the kinetically decoupled result to assess the impact on $N_{\rm eff}$.
	
	The right panel of Fig. \ref{neff2} shows two contours satisfying two $\Delta N_{\rm eff}$ values in the $M_{Z'} ~vs.~ M_\psi$ plane. The two dotted lines represent two resonance benchmark points satisfying DM relic in the same plane.

	As discussed earlier, we have an additional contribution of ${\Delta N_{\rm eff}}=0.0466$ from the light $\nu_{eR}$ decoupling. Now, from the decay and annihilation of $Z'$ and DM particle, we have new contributions to ${\Delta N_{\rm eff}}$ value. The total effective ${\Delta N_{\rm eff}}$ value will be the sum of the contributions arising from all the sources.

	\section{Conclusion}\label{sec4}
	We have studied a Dirac seesaw framework with the local $U(1)_{L_\mu-L_\tau}$ gauged symmetry, where the new gauge boson $Z'$ and the corresponding gauge coupling $g_{\mu\tau}$ play key roles throughout our study. In a chronological manner, we first set up a workable Dirac neutrino mass model. Then we choose our desired parameter space from the recent experiments, constraining the $M_{Z'}-g_{\mu\tau}$ region. The region under study is within the range of $M_{Z'}=[1,200]$ MeV with a fixed coupling strength of $g_{\mu\tau}=3.8\times10^{-4}$. Among the model ingredients, the heavy particles decay and leave the thermal bath along with the $\nu_{eR}$, while the other two light right-handed Dirac neutrinos ($\nu_{(\mu,\tau) R}$) never enter the thermal bath due to the choice of specific gauge charges.
	
	In the DM study, we divide the sector into two parts by the $Z'$ mass (at 100 MeV), which is consistent with the latest $(g-2)_\mu$ result. Dark matter parameters consistent with the current best-fit value of the relic density ($\Omega_{\rm DM}h^2$) are obtained in both regions, primarily focusing near the resonance region ($M_{\rm DM}\simeq M_{Z'}/2$). We also show that kinetic decoupling occurs sufficiently early in the resonant regime, remaining above the neutrino decoupling temperature throughout the benchmark parameter space considered. {Consequently, the standard entropy-transfer treatment employed in the computation of cosmological observables gets modified. However, we find that the effect of early kinetic decoupling on the relic abundance is negligible. Therefore, we retain the standard freeze-out treatment for the relic-density calculation in this analysis. } By requiring kinetic decoupling to occur before neutrino decoupling, we obtain a consistency condition of approximately $M_{Z'} \gtrsim 20{\rm MeV}$. In the resonant regime relevant for relic-density production, where $M_{\rm DM}\simeq M_{Z'}/2$, this translates to $M_{\rm DM}\gtrsim 10{\rm MeV}$. Although this condition is not stronger than the existing BBN and Planck bound of $M_{\rm DM}\gtrsim 10.9{\rm MeV}$, it provides an independent consistency check of the thermal history assumed in our analysis and further supports the applicability of the entropy-transfer treatment used in the calculation of $\Delta N_{\rm eff}$.

	The additional contributions to $N_{\rm eff}$ are calculated via all possible processes contributing to the radiation density. { Among the three RH neutrinos, we restrict the gauge charge of $\nu_{\mu R}$ and $\nu_{\tau R}$ in such a fashion that they never enter the thermal bath.} The $\nu_eR$ will leave the thermal bath at a very early time (before the EWSB) and it will give a fixed contribution of $\Delta N_{\rm eff}=0.0466$. The final value of $\Delta N_{\rm eff}$ will be the combined contributions of all possibilities. {We identify the regions in the $M_{Z'}-M_\psi$ parameter space compatible with the $\Delta N_{\rm eff}$ constraints, which come from the production of SM neutrinos.} To keep on the safe side, we check the spin-independent $DM-electron$ scattering cross-section, as such, light DM masses will be sensitive only to them, which remains suppressed due to the factor $m_e/M_{Z'}<<1$. 
	
	\section*{Acknowledgements}
	The author thanks Debasish Borah (IIT-Guwahati) for suggesting the idea of this work. Various aspects of this research were discussed with Arunansu Sil (IIT-Guwahati) and Debasish Borah (IIT-Guwahati), and the author is grateful to them for insightful discussions. The author also sincerely thanks Satyabrata Mahapatra (IIT-Goa) for carefully reading \& reviewing the manuscript and providing valuable comments and suggestions that improved the presentation of this work. The author acknowledges Devabrat Mahanta (Pragjyotish College) and Nayan Das (IIT-Guwahati) for discussions on various aspects of this work.
	
	\appendix
	{\section{Consistency check for the relic-density calculation with early kinetic decoupling}\label{apendix1}
		Due to the Breit-Wigner resonance enhancement, the kinetic decoupling happens prior to the chemical freeze-out. This situation makes the DM follow a different temperature than the SM bath temperature. Hence the DM temperature evolves independently. The standard relic-density calculation assumes kinetic equilibrium, $(T_\psi=T)$, during chemical freeze-out. This assumption can be relaxed by retaining the second moment of the dark-matter phase-space distribution. Following the numerical treatment by \cite{Binder:2017rgn}, we calculated the relic abundance of the DM for the standard scenario and for the coupled system. The numerical treatment is briefly resented here for our case as follows. 
		
		The coupled Boltzmann equations can be written as
		\begin{eqnarray}
			\frac{Y'}{Y}=
			\frac{sY}{x\widetilde H}
			\left[
			\frac{Y_{\rm eq}^2}{Y^2}
			\left\langle\sigma v\right\rangle_T-\left\langle\sigma v\right\rangle_{T_\psi}
			\right],
		\end{eqnarray}
		and
		\begin{eqnarray}
			\begin{aligned}
				\frac{y'}{y}
				={}&
				\frac{1}{x\widetilde H}
				\left\langle C_{\rm el}\right\rangle_2+
				\frac{sY}{x\widetilde H}
				\left[
				\left\langle\sigma v\right\rangle_{T_\psi}-\left\langle\sigma v\right\rangle_{2,T_\psi}
				\right]
				\
				+
				\frac{sY}{x\widetilde H}
				\frac{Y_{\rm eq}^2}{Y^2}
				\left[
				\frac{y_{\rm eq}}{y}
				\left\langle\sigma v\right\rangle_{2,T}-\left\langle\sigma v\right\rangle_T
				\right]
				+
				2(1-w)\frac{H}{x\widetilde H}
			\end{aligned}
		\end{eqnarray}
		
		Here, a prime denotes differentiation with respect to $x$ with 
		$\widetilde H=
		\frac{H}{1+\widetilde g},$ and $
		\widetilde g=
		\frac{1}{3}
		\frac{T}{g_{*s}}
		\frac{d g_{*s}}{dT}$. The notations and definitions are kept consistent with the work of \cite{Binder:2017rgn}.
		
		The equilibrium value of the second-moment variable is $y_{\rm eq}=\frac{m_\psi T}{s^{2/3}}$, such that $(T_\psi=T)$ when kinetic equilibrium is maintained. The quantity $\langle C_{\rm el}\rangle_2$ describes the effect of elastic scattering on the second moment. In the Fokker-Planck approximation it can be expressed in terms of the momentum-transfer rate $\gamma(T)$, schematically as
		$$
		\left\langle C_{\rm el}\right\rangle_2
		\simeq
		\gamma(T)w(T_\psi)
		\left(
		\frac{y_{\rm eq}}{y}-1
		\right).
		$$
		The additional annihilation average $\langle\sigma v\rangle_{2,T_\psi}$ is the temperature-weighted thermal average entering the second moment equation. The quantity $w(T_\psi)$ accounts for the departure from the strictly non-relativistic limit. These equations retain the information about the dark-matter velocity distribution that is lost in the conventional number-density equation.
		
		We have solved the above equations numerically for the benchmark parameter point used in our relic-density analysis and compared the result with the conventional single Boltzmann equation. The coupled solution confirms that the dark-matter temperature starts to deviate from the bath temperature after kinetic decoupling $T_\psi<T$ as expected. For the benchmark point used in our analysis, we find that the asymptotic relic abundance obtained from the coupled treatment differs from that obtained using the standard single Boltzmann equation by only
		\begin{eqnarray}
			\left|\frac{\Omega h^2_{\rm coupled}-\Omega h^2_{\rm standard}}
			{\Omega h^2_{\rm standard}}\right|
			\simeq 0.002162,
		\end{eqnarray}
		corresponding to a correction of approximately (0.22\%). Therefore, while kinetic equilibrium is indeed violated after early kinetic decoupling, the resulting modification of the thermally averaged annihilation cross section does not lead to a significant change in the integrated annihilation history or the final relic abundance for the benchmark point considered in this work.
		
		Thus, while kinetic equilibrium is not maintained throughout the entire period following early kinetic decoupling, the resulting modification of the annihilation history is too small to produce an appreciable change in the final relic abundance for the benchmark parameter point considered here. Consequently, the relic-density results presented in Fig. \ref{resonant} remain unchanged after the coupled treatment.
		
		We stress that this agreement is a numerical result for the parameter region considered in this work and should not be interpreted as a general statement that early KD always produces negligible changes in the DM relic calculation. Rather, the coupled calculation provides an explicit validation that the standard relic-density treatment is sufficiently accurate for the benchmark scenario studied here.}
	
	\bibliographystyle{JHEP}
	\bibliography{ref,ref1,ref2, Scottref}
\end{document}